\documentclass[12pt]{article}
\usepackage{amsmath}

\begin{document}

\author{Giovanni Montani$^{(1)}$\thanks{
E-mail: montani@icra.it}, Remo Ruffini$^{(1)}$\thanks{
E-mail: ruffini@icra.it} \ and Roustam Zalaletdinov$^{(1,2,3)}$\thanks{
E-mail: zala@icra.it} \\ [5mm]
\emph{$^{(1)}$ICRA, Departamento di Fisica, Universit\'a di Roma ``La
Sapienza"} \\
\emph{P.le Aldo Moro 5, Roma 00185, Italia} \\
[2mm] \emph{$^{(2)}$Department of Mathematics and Statistics, Dalhousie
University} \\
\emph{Chase Building, Halifax, Nova Scotia, Canada B3H 3J5} \\
[2mm] \emph{$^{(3)}$Department of Theoretical Physics, Institute of Nuclear
Physics} \\
\emph{Uzbek Academy of Sciences, Tashkent 702132, Uzbekistan, CIS}}
\title{{\LARGE {\textbf{The Gravitational Polarization in General
Relativity: Solution to Szekeres' Model of Quadrupole Polarization }}}}
\date{}
\maketitle

\begin{abstract}
A model for the static weak-field macroscopic medium is analyzed and the
equation for the macroscopic gravitational potential is derived. This is a
biharmonic equation which is a non-trivial generalization of the Poisson
equation of Newtonian gravity. In case of the strong gravitational
quadrupole polarization it essentially holds inside a macroscopic matter
source. Outside the source the gravitational potential fades away
exponentially. The equation is equivalent to a system of the Poisson
equation and the nonhomogeneous modified Helmholtz equations. The general
solution to this system is obtained by using Green's function method and it
does not have a limit to Newtonian gravity. In case of the insignificant
gravitational quadrupole polarization the equation for macroscopic
gravitational potential becomes the Poisson equation with the matter density
renormalized by the factor including the value of the quadrupole
gravitational polarization of the source. The general solution to this
equation obtained by using Green's function method has a limit to Newtonian
gravity.
\end{abstract}

\section{Introduction}
\label{problem}
\noindent
General relativity as a classical theory of gravity is known to have some
remarkable analogies with classical Maxwell's macroscopic theory of
electromagnetism (see, for example, \cite{MTW:1973} and references therein).
The physical motivation and intuition for many of the problems posed in
general relativity originate therefore in their electromagnetic analogies
where physics and formalism are much easier to deal with. One of such
problems of the primary importance is that of gravitational waves which has
been inspired and put forward mostly owing to our deep understanding of the
structure and physics of electromagnetic waves. Some important issues in the
physics of gravitational waves in general relativity, however, remain
obscure. The questions whether or not gravitational waves undergo the
refraction in a gravitating macroscopic (continuous) media, the speed of a
gravitational wave changes (i.e. slows down) in a material medium, the
phenomenon of gravitational polarization exists, have been not approached
and even properly posed as yet.

To determine the structure of the macroscopic energy-momentum
tensor resulting from averaging out a microscopic matter source,
the problem of construction of a continuous (macroscopic) matter
model for a given point-like (microscopic) matter distribution in
general relativity has been formulated in \cite{MRZ:2001a}. The
existing approaches have considered and a physical analogy with
the similar problem in classical macroscopic electrodynamics has
been pointed out. The procedure due to Szekeres \cite {Szek:1971}
in the linearized general relativity on Minkowski background
space-time to construct a tensor of gravitational quadruple
polarization by applying Kaufman's method of molecular moments
\cite{Kauf:1962} for derivation of the polarization tensor in
macroscopic electrodynamics and to derive an averaged field
operator by utilizing an analogy between the linearized Bianchi
identities and Maxwell equations, has been analyzed. The approach
of Szekeres to construct a tensor of the gravitational quadruple
polarization is based on the following assumptions: (a) the
linearized theory of gravity on Minkowski background space-time;
(b) the linearized field equations are taken as the linearized
Bianchi identity to employ an analogy between gravitation and
electromagnetism; (c) the covariant method of molecular moments of
Kaufman is applied to construct a tensor of quadruple
gravitational polarization. The procedure is shown to possess some
inconsistencies, in particular, (1) it has only provided the terms
linear in perturbations for the averaged field operator which do
not contribute into the dynamics of the averaged field, and (2)
the analogy between electromagnetism and gravitation does break
upon averaging. A macroscopic gravity approach in the perturbation
theory up to the second order on a particular background
space-time taken to be a smooth weak gravitational field has been
applied to write down a system of macroscopic field equations
\cite{MRZ:2001a}, \cite{MRZ:2001b}, \cite{MRZ:2001c}: Isaacson's
equations \cite{Isaa:1968a} with a source incorporating the
quadruple gravitational polarization tensor, Isaacson's
energy-momentum tensor of gravitational waves \cite{Isaa:1968b}
and energy-momentum tensor of gravitational molecules and
corresponding equations of motion. The system of equations is
shown to be underdetermined. A suitable set of material relations
which relate all the tensors has been proposed \cite{MRZ:2001a},
\cite{MRZ:2001b}, \cite{MRZ:2001c}, so that the full system of the
field equations and the material relations become determined.

In this paper the system of equations is used to find a solution to the
Szekeres model of the gravitational quadrupole polarization. A model for the
static weak-field macroscopic medium is analyzed and the equation for the
macroscopic gravitational potential is derived. This is a biharmonic
equation which generalizes the Poisson equation of Newtonian gravity. In
case of the strong gravitational quadrupole polarization it essentially
holds inside a macroscopic matter source. Outside the source the
gravitational potential fades away exponentially. The equation is equivalent
to a system of the Poisson equation and the nonhomogeneous modified
Helmholtz equations. The general solution to this system is obtained by
using Green's function method. This solution does not have a limit to
Newtonian gravity. In case of the insignificant gravitational quadrupole
polarization the equation for macroscopic gravitational potential becomes
the Poisson equation with the matter density renormalized by the factor
including the value of the quadrupole gravitational polarization of the
source. The general solution to this equation has been obtained by using
Green's function method and it has a limit to Newtonian gravity.

\section{Macroscopic Description of Gravity}
\label{*mdg} \noindent To consider the problem of macroscopic
description of gravitation, an approach of macroscopic gravity has
been proposed earlier in \cite{Zala:1992}, \cite{Mars-Zala:1997}
(see \cite{Zala:1997}, \cite{Zala:1998}, \cite{Kras:1997} for
discussion of the problem and references therein, \cite
{Tava-Zala:1998} for discussion of the physical status of general
relativity as either a microscopic or macroscopic theory of
gravity). A covariant space-time volume averaging procedure for
tensor fields \cite{Zala:1992}, \cite{Mars-Zala:1997},
\cite{Zala:1993} has been defined and proved to exist on arbitrary
Riemannian space-times with well-defined properties of the
averages. Upon utilizing the averaging scheme, the macroscopic
gravity approach has shown that (\emph{i}) averaging out Cartan's
structure equations brings about the structure equations for the
averaged (macroscopic) non-Riemannian geometry and the definition
and the properties of the correlation tensor, (\emph{ii}) the
averaged Einstein's equations become then the macroscopic field
equations and they must be supplemented by a set of differential
equations for the correlation tensor, (\emph{iii}) it is always
possible to extract the field operator of the same form as that of
Einstein's equations for the Riemannian macroscopic metric tensor
and its Ricci tensor while all other non-Riemannian correlation
terms go to the right-hand side of the averaged Einstein's
equations to give geometric correction to the averaged
(macroscopic) energy-momentum tensor. It is been also shown
\cite{Zala:1997}, \cite{Zala:1998}, \cite{Zala:1996b} that only in
case of neglecting all correlations of the gravitational field the
averaged equations becomes the macroscopic Einstein equations for
the Riemannian macroscopic metric tensor and its Ricci tensor with
a continuous matter distribution. This result reveals the physical
status of using the so-called standard procedure in cosmology when
one claims that the Einstein equations preserve their form after
substitution of a discrete matter model by a continuous one.

In using the Einstein equations for a matter distribution in the
form of a set of point-like mass constituents, there is a problem
of adequate application, or validity, of the Einstein equations
when such a matter distribution is substituted by a continuous
matter distribution while the field operator in the left-hand side
of the equations is kept unchanged. This problem as it stands in
cosmology\footnote{For discussion on the other physical
settings on general relativity facing the same problem see
\cite{Tava-Zala:1998}.} is called the averaging problem
\cite{Zala:1992}, \cite{Shir-Fish:1962}, \cite{Scia:1971},
\cite{Elli:1984}, \cite{Zoto-Stoe:1992}. Indeed, let us consider
the Einstein equations in the mixed form\footnote{The mixed
form is preferable here for the reason that it contains only
products of metric by curvature. On contrary, the covariant or
contravariant forms of the Einstein equations have triple products
of metric by metric by curvature.}
\begin{equation}
g^{\alpha \epsilon }r_{\epsilon \beta }-\frac{1}{2}\delta _{\beta }^{\alpha
}g^{\mu \nu }r_{\mu \nu }=-\kappa t_{\beta }^{\alpha \mathrm{(discrete)}}
\label{EE}
\end{equation}
with
\begin{equation}
t_{\beta }^{\alpha \mathrm{(discrete)}}(x)=\sum_{i}^{{}}{t_{(i)}}_{\beta
}^{\alpha }[x-z_{(i)}(\tau _{(i)})]  \label{dmd}
\end{equation}
where ${t_{(i)}}_{\beta }^{\alpha }$ is a energy-momentum tensor
for a point-like mass moving along its world line $z^{\mu
}=z_{(i)}^{\mu }(\tau _{(i)})$ parameterized by $\tau _{(i)}$ and
$i$ counts for the matter particles in the distribution
(\ref{dmd}). Changing the discrete matter distribution to a
continuous (hydrodynamic) one in the right-hand side of
(\ref{EE}), which is the standard approach in cosmology
\cite{Zala:1992}, \cite{Shir-Fish:1962}, \cite{Scia:1971},
\cite{Elli:1984}, \cite{Zoto-Stoe:1992} made phenomenologically on
the basis of assumption about the uniformity and isotropicity of
distribution of galaxies, or cluster of galaxies, throughout the
whole Universe, means an implicit averaging denoted here by
$\langle \cdot \rangle $
\begin{equation}
t_{\beta }^{\alpha \mathrm{(discrete)}}(x)\rightarrow T_{\beta }^{\alpha
\mathrm{(hydro)}}(x)=\left\langle \sum_{i}^{{}}{t_{(i)}}_{\beta }^{\alpha
}[x-z_{(i)}(\tau _{(i)})]\right\rangle \ .  \label{aver-dmd}
\end{equation}
Given a covariant averaging procedure $\langle \cdot \rangle $ for
tensors on space-time, the averaging out of (\ref{EE}) with taking
into account (\ref{aver-dmd}) brings
\begin{equation}
\langle g^{\alpha \epsilon }r_{\epsilon \beta }\rangle -\frac{1}{2}\delta
_{\beta }^{\alpha }\langle g^{\mu \nu }r_{\mu \nu }\rangle =-\kappa T_{\beta
}^{\alpha \mathrm{(hydro)}}\ .  \label{averEE:1}
\end{equation}
An important point regarding the averaged equations (\ref{averEE:1}) is that
in this form they are just algebraic relations between components of the
smoothed hydrodynamic energy-momentum tensor and the average products of the
metric tensor by the Ricci tensor $\langle g^{\alpha \epsilon }r_{\epsilon
\beta }\rangle $ and cannot therefore be taken as field equations. By
splitting the products out as $\langle g^{\alpha \epsilon }r_{\epsilon \beta
}\rangle =\langle g^{\alpha \epsilon }\rangle \langle r_{\epsilon \beta
}\rangle +C_{\beta }^{\alpha }$ where $C_{\beta }^{\alpha }$ is a
correlation tensor, the averaged equations (\ref{averEE:1}) become
\begin{equation}
\langle g^{\alpha \epsilon }\rangle \langle r_{\epsilon \beta
}\rangle - \frac{1}{2}\delta _{\beta }^{\alpha }\langle g^{\mu \nu
}\rangle \langle r_{\mu \nu }\rangle =-\kappa T_{\beta }^{\alpha
\mathrm{(hydro)}}-C_{\beta }^{\alpha }+\frac{1}{2}\delta _{\beta
}^{\alpha }C_{\epsilon }^{\epsilon }. \label{averEE:2}
\end{equation}
Here $\langle g^{\alpha \beta }\rangle $ and $\langle r_{\alpha
\beta }\rangle $ denote the averaged inverse metric and the Ricci
tensors which are supposed to describe the gravitational field due
to the matter distribution $T_{\beta }^{\alpha \mathrm{(hydro)}}$.
A simple important observation \cite{Zala:1997}, \cite{Zala:1998}
now is that the averaged Einstein equations (\ref{averEE:2}) are
still not \textquotedblleft real" field equations - just a
definition of the correlation tensor $C_{\beta }^{\alpha }$ as a
difference between (\ref{averEE:1}) and (\ref{averEE:2}). The
origin of this fundamental fact is that the average of the
non-linear operator of (\ref{EE}) on the metric tensor $g_{\rho
\sigma }$ is not equal in general\footnote{It should noted
that the inequality (\ref{aver:oper}) has been observed in all
possible averaging settings, for example, for a volume space-time
averaging in \cite{Zala:1992}, \cite{Shir-Fish:1962}, in the
framework of a kinetic approach in \cite{Yodz:1971} and for a
statistical ensemble averaging in \cite{Igna:1978}. Relations
between different averaging procedures are discussed in
\cite{Zala:1997}, \cite{Zala:1998}.} to an operator of the same
form on the average metric $\langle g_{\rho \sigma }\rangle $:
\begin{equation}
\left\langle \left( g^{\alpha \epsilon }r_{\epsilon \beta
}-\frac{1}{2} \delta _{\beta }^{\alpha }g^{\mu \nu }r_{\mu \nu
}\right) [g_{\rho \sigma }]\right\rangle \neq \left( \langle
g^{\alpha \epsilon }\rangle \langle r_{\epsilon \beta }\rangle
-\frac{1}{2}\delta _{\beta }^{\alpha }\langle g^{\mu \nu }\rangle
\langle r_{\mu \nu }\rangle \right) [\langle g_{\rho \sigma
}\rangle ]\ .  \label{aver:oper}
\end{equation}
In order to return them the status of the field equations one must define
the object $C_{\beta }^{\alpha }$ and find its properties using information
outside the Einstein equations.

To resolve the averaging problem, and to consider it in a broader
context as the problem of macroscopic description of gravitation,
the approach of macroscopic gravity has been proposed
\cite{Zala:1992}, \cite{Mars-Zala:1997}, \cite{Zala:1997},
\cite{Zala:1998}, \cite{Kras:1997}, \cite{Zala:1993},
\cite{Zala:1996b}, \cite{Zala:1996} (see \cite{Zala:1997},
\cite{Kras:1997} for discussion of the problem and references
therein,~\cite{Tava-Zala:1998} for discussion of the physical
status of general relativity as either a microscopic or
macroscopic theory of gravity). A covariant space-time volume
averaging procedure for tensor fields \cite{Zala:1992},
\cite{Mars-Zala:1997}, \cite{Zala:1993}, has been defined and
proved to exist on arbitrary Riemannian space-times with
well-defined properties of the averages. Upon utilizing the
averaging scheme, the macroscopic gravity approach has shown that
(\emph{i}) averaging out Cartan's structure equations brings about
the structure equations for the averaged (macroscopic)
non-Riemannian geometry and the definition and the properties of
the correlation tensor $C_{\beta }^{\alpha }$, (\emph{ii}) the
averaged Einstein's equations (\ref{averEE:2}) become then the
macroscopic field equations and they must be supplemented by a set
of differential equations for the correlation tensor, (\emph{iii})
it is always possible to extract the field operator of the form
(\ref{EE}) for the Riemannian macroscopic metric tensor $G_{\mu
\nu }$ and its Ricci tensor $M_{\mu \nu }$ with all other
non-Riemannian correlation terms going to the right-hand side of
(\ref{averEE:2}) to give geometric correction to the averaged
energy-momentum tensor $T_{\beta }^{\alpha \mathrm{(hydro)}}$. It
is been also shown \cite{Zala:1997}, \cite{Zala:1998},
\cite{Zala:1996b} that only in case of neglecting all correlations
of the gravitational field the averaged equations (\ref{averEE:2})
becomes the macroscopic Einstein equations with a continuous
matter distribution
\begin{equation}
G^{\alpha \epsilon }M_{\epsilon \beta }-\frac{1}{2}\delta _{\beta }^{\alpha
}G^{\mu \nu }M_{\mu \nu }=-\kappa T_{\beta }^{\alpha \mathrm{(hydro)}},
\label{macroEE}
\end{equation}
which reveals the physical status of using the standard procedure
in cosmology of claiming (\ref{averEE:2}) to be the Einstein
equations (\ref{macroEE}) after substitution of the matter model
(\ref{aver-dmd}). The physical meaning, dynamical role and
magnitude of the gravitational correlations must be elucidated in
various physical settings. There is some evidence that they cannot
be negligible for cosmological evolution (see, for example
\cite{Bild-Futa:1991} for an estimation of the age of Universe in
a second order perturbation approach).

\section{Macroscopic media in general relativity}
\label{media} \noindent Derivation of the macroscopic (averaged)
Maxwell field operator in macroscopic electrodynamics is easily
accomplished due its linear field structure and the main problem
consists in the construction of models of macroscopic
electromagnetic media (for example, diamagnetics, magnetics,
waveguides, etc.) \cite{Lore:1916}, \cite{Pano-Phil:1962},
\cite{deGr-Sutt:1972}, \cite{Jack:1975}, which relates to the
structure of the averaged current. In general relativity the
problem of construction of macroscopic gravitating medium models
is hardly elaborated due to the following reasons: (a) existing
mathematical and physical difficulties in establishing the form of
the averaged (macroscopic) operator in (\ref{averEE:2}) for the
field equations of macroscopic gravity recedes the interest in
development of macroscopic gravitating media; (b) posing on its
own more or less realistic problem with a discrete matter creates
mathematical and physical problems due to nonlinearity and
non-trivial geometry of gravitation, to mention, for example, the
$N$-body problem, the problem of statistical description of
gravity, etc. and (c) being relied on physically motivated
phenomenological arguments (uniformity, isotropy, staticity, etc.)
most applications of general relativity deal with \emph{effective}
continuous media if even a starting physical model is discrete in
its nature like in cosmology (see Section \ref{*mdg}) or in
description of extended bodies in general relativity (see
\cite{Tava-Zala:1998} for discussion of the physical status of
general relativity).

The kinetic approach in physics is known to provide a general
scheme for introduction of characteristics of continuous media
with a \emph{known} distribution function of a discrete
configuration. But the advantages of such generality are often
greatly weakened in particular applications by difficulties of
solving the Boltzmann equation to find a distribution function of
interest. This applies to a great extent to general relativity
where despite the formulation of the general relativistic
Boltzmann equation \cite{Cher:1962}, \cite{Isra:1972} the kinetic
approach still remains useful for general definitions and
considerations rather than being a working tool (see, for example,
\cite{Yodz:1971}) to derive a specific model of a macroscopic
medium.

In case of the macroscopic electrodynamics together with the volume
space-time averaging on Minkowski space-time the formalism of statistical
distribution functions has been utilized (see \cite{deGr-Sutt:1972} and
references therein) and it is of importance for the mathematically
well-posed derivation of the macroscopic theory and the general structure of
averaged current starting from microscopic electrodynamics of point-like
moving charges. Further application of the macroscopic theory requires
usually mainly phenomenological considerations to establish material
relations between macroscopic average fields and induction field necessary
to make an overdetermined system of macroscopic equations determined. A
correct derivation of material relations is known to require \cite{deGr:1969}
averaging the microscopic equations with a given microscopic matter model
during accomplishing an averaging of the microscopic field equations. Though
it is the only self-consistent way, the elaboration of such kind of approach
still remains a challenge even for simple physical settings.

On the other hand, volume (space, time, space-time) averaging
procedures maintain their importance, direct physical meaning in
application to macroscopic settings and their extreme clearness
and descriptiveness. A volume averaging is also known to be
unavoidable in all macroscopic settings (including statistical
approaches) \cite{Jack:1975}, \cite{deGr:1969}, \cite{Russ:1970},
\cite{Robi:1973} and space-time averages of physical fields are
known to have the physical meaning as directly measurable
quantities \cite{Bohr-Rose:1933}, \cite{DeWi:1962} (for discussion
see \cite{Zala:1997}, \cite{Zala:1998} and references therein).
That greatly motivates and supports interest in applying
approaches with various averaging schemes in physics despite
corresponding (mostly mathematical, not physical) difficulties in
the rigorous formulation of averaging procedures.

The paper aims to approach the problem of construction of gravitating
macroscopic media in general relativity by using an appropriate space-time
averaging scheme.

\section{Szekeres' gravitational polarization tensor}
\noindent
The approach of Szekeres~\cite{Szek:1971} to construct a tensor of
gravitational quadruple polarization is based on the following assumptions:
(a) the linearized theory of gravity on Minkowski background; (b) the
linearized field equations are taken as the linearized Bianchi identity to
employ an analogy between gravitation and electromagnetism; (c) the
covariant method of molecular moments of Kaufman is applied to construct a
tensor of quadruple gravitational polarization.

The equations under consideration are the contracted Bianchi identities
\begin{equation}  \label{bianchi}
C_{\mu \nu \rho \sigma}{}^{; \sigma} = \kappa J_{\mu \nu \rho}
\end{equation}
where $C_{\mu \nu \rho \sigma}$ is the Weyl tensor interpreted as free
gravitational field, $J_{\mu \nu \rho}$
is a kind of ``matter current" for the energy-momentum tensor $t_{\mu \nu}$
\begin{equation}  \label{matter-current}
J_{\mu \nu \rho} = J_{[ \mu \nu ] \rho} = - (t_{\rho [ \mu ; \nu ]} -
\frac{1}{3} g_{\rho [ \mu} t_{, \nu ]}) ,
\end{equation}
\begin{equation}  \label{matter-current-conserv}
J_{\mu \nu \rho}{}^{; \rho} = 0 .
\end{equation}
equations (\ref{bianchi}) are analogous the Maxwell equations
\begin{equation}  \label{maxwell}
f_{\mu \nu }{}^{; \nu} = \frac{4 \pi}{c} j_{\mu}
\end{equation}
with (\ref{matter-current-conserv}) being comparable with the conservation
of the electromagnetic current $j_{\mu}$
\begin{equation}  \label{em-current-conserv}
j_{\mu}{}^{; \mu} = 0 .
\end{equation}

Let us consider a number of particles labelled by $i$ and having masses $m_{i}
$ and which are moving in there own effective gravitational field along
world lines $z_{i}^{\mu }(\tau _{i})$. A physical parameter which
characterizes such a distribution is a typical characteristic distance $l$
between neighboring particles. Then the corresponding microscopic
energy-momentum tensor has the form
\begin{equation}
t^{\mathrm{(micro)}\mu \nu }(x)=c^{-1}\sum_{i}\int m_{i}\frac{dz_{i}^{\mu }}
{d\tau _{i}}\frac{dz_{i}^{\nu }}{d\tau _{i}}\delta ^{4}[x-z_{i}^{\mu }(\tau
_{i})]d\tau _{i}.  \label{micro}
\end{equation}
Assume now that due to gravitation the particles form into groups, a kind of
gravitational molecules, which will be labelled by index $a$. From the
physical point of view that means the presence of another parameter which is
a characteristic size (diameter) $L$ of such a molecule with $L\gg l$. It is
a longwave \emph{macroscopic} parameter and its presence in a microscopic
system will be defining the dynamics of the system on the distances of order
of $L$. The microscopic energy-momentum tensor (\ref{micro}) becomes now
\begin{equation}
t^{\mathrm{(molec)}\mu \nu }(x)=c^{-1}\sum_{a}\sum_{i}^{ina}\int m_{i}
\frac{d\tau _{a}}{d\tau _{i}}\frac{dz_{i}^{\mu }}{d\tau _{a}}\frac{dz_{i}^{\nu }}
{d\tau _{a}}\delta ^{4}[x-y_{a}^{\mu }(\tau _{a})-s_{i}(\tau _{a})]d\tau _{a}
\label{molec}
\end{equation}
where $y_{a}^{\mu }(\tau _{a})$ is a world line of the $a$-th molecule
center of mass \cite{Szek:1971} and $s_{i}^{\mu }(\tau _{a})=z_{i}^{\mu
}(\tau _{i})-y_{a}^{\mu }(\tau _{a})$ is a vector connecting the $i$-th
particle with the center of mass of the molecule including this particle.
Let us apply now the method of molecular moments of Kaufman \cite{Kauf:1962}
to represent (\ref{molec}) as a series expansion in powers of $s_{i}^{\mu }$
under assumption that the effective gravitational field which is created by
moving gravitational molecules is a weak field and the perturbations of the
gravitational field due to relative oscillations of gravitating particles in
molecules are small compared with the mean effective field. After averaging
out over a typical size of the gravitational molecule (for an averaging
procedure see \cite{Zala:1992}, \cite{Zala:1993}) one
gets\footnote{No explicit averaging procedure had been used in
\cite{Szek:1971}
and averaged relations and equations were being written rather on
the basis of heuristic considerations than a rigorous analysis.}
in accordance with the Szekeres procedure
\begin{equation}
\langle t^{\mathrm{(molec)}\mu \nu }\rangle =T^{\mathrm{(free)}\mu \nu
}+D^{\mu \nu \rho }{}_{,\rho }+Q^{\mu \nu \rho \sigma }{}_{,\rho \sigma }
\label{av-molec}
\end{equation}
where $T^{\mathrm{(free)}\mu \nu }$ is the energy-momentum tensor of
molecules, which has the form similar to (\ref{micro}) with substitution $i$
by $a$, $D^{\mu \nu \rho }$ is the tensor of gravitational dipole
polarization that can be incorporated into the quadrupole term, and $Q^{\mu
\nu \rho \sigma }$ the tensor of gravitational quadrupole polarization
\begin{equation}
Q^{\mu \nu \rho \sigma }=c^{-1}\langle \sum_{a}\int q_{a}^{\mu \nu \rho
\sigma }\delta ^{4}(x-y_{a})d\tau _{a}\rangle ,  \label{polar}
\end{equation}
which has the symmetries of the Riemann tensor. The expression for the
covariant gravitational quadrupole moment $q_{a}^{\mu \nu \rho \sigma }$ is
defined as
\begin{eqnarray}
q_{a}^{\mu \nu \rho \sigma } & = & g_{a}^{\mu \nu }u_{a}^{\rho }u_{a}^{\sigma
}-g_{a}^{\rho \nu }u_{a}^{\mu }u_{a}^{\sigma }-g_{a}^{\mu \sigma
}u_{a}^{\rho }u_{a}^{\nu }+g_{a}^{\rho \sigma }u_{a}^{\mu }u_{a}^{\nu
}+ \nonumber \\
& & u_{a}^{\mu }h_{a}^{\rho \nu \sigma }-u_{a}^{\rho }h_{a}^{\mu \nu \sigma
}+u_{a}^{\nu }h_{a}^{\sigma \mu \rho }-u_{a}^{\sigma }h_{a}^{\nu \mu \rho
}+k_{a}^{\mu \nu \rho \sigma },  \label{moment}
\end{eqnarray}
where
\begin{equation}
g_{a}^{\mu \nu }=\sum_{i}m_{i}\frac{d\tau _{a}}{d\tau _{i}}s_{i}^{\mu
}s_{i}^{\nu },
\end{equation}
\begin{equation}
h_{a}^{\mu \nu \rho }=\frac{2}{3}\sum_{i}m_{i}\frac{d\tau _{a}}{d\tau _{i}}
s_{i}^{\mu }\left( \frac{ds_{i}^{\nu }}{d\tau _{a}}s_{i}^{\rho }-\frac{
ds_{i}^{\rho }}{d\tau _{a}}s_{i}^{\nu }\right) ,
\end{equation}
\begin{equation}
k_{a}^{\mu \nu \rho \sigma }=\frac{2}{3}\sum_{i}m_{i}\frac{d\tau _{a}}{d\tau
_{i}}\left( \frac{ds_{i}^{\mu }}{d\tau _{a}}s_{i}^{\rho }\frac{ds_{i}^{\nu }
}{d\tau _{a}}s_{i}^{\sigma }-\frac{ds_{i}^{\mu }}{d\tau _{i}}s_{i}^{\rho
}s_{i}^{\nu }\frac{ds_{i}^{\sigma }}{d\tau _{a}}\right) .
\end{equation}
Upon averaging (\ref{bianchi}) over the typical size of gravitational
molecule the following equations were obtained:
\begin{equation}
{\langle C_{\mu \nu \rho \sigma }{}\rangle }^{,\sigma }=\kappa \langle
J_{\mu \nu \rho }^{\mathrm{(micro)}}\rangle   \label{av-bianchi}
\end{equation}
where
\begin{equation}
\langle J_{\mu \nu \rho }^{\mathrm{(micro)}}\rangle =-\langle t_{\rho
\lbrack \mu }^{\mathrm{(micro)}}\rangle _{,\nu ]}+\frac{1}{3}\eta _{\rho
\lbrack \mu }\langle t^{\mathrm{(micro)}}\rangle _{,\nu ]},
\label{av-matter-current}
\end{equation}
or
\begin{equation}
P_{\mu \nu \rho \sigma }=\frac{1}{2}(-{Q_{\rho \sigma \epsilon \lbrack \mu }}
^{,\epsilon }-\frac{1}{3}\eta _{\rho \lbrack \mu }Q^{\gamma }{}_{\sigma
\gamma \epsilon }{}^{,\epsilon })_{,\nu ]}  \label{polarization}
\end{equation}
and
\begin{equation}
\langle J_{\mu \nu \rho }^{\mathrm{(micro)}}\rangle =J_{\mu \nu \rho }^{
\mathrm{(free)}}-P_{\mu \nu \rho \sigma }{}^{,\sigma }.
\label{av-matter-current-2}
\end{equation}
The expression (\ref{av-matter-current-2}) is analogous to the expression
for the averaged electromagnetic current $\langle j^{\mathrm{(micro)\mu }
}\rangle $ for a bunch of charged particles moving along their world lines
in the effective electromagnetic field in accordance with the microscopic
equation (\ref{maxwell}) with $j^{\mathrm{(micro)\mu }}$ when particles are
grouped into molecules \cite{Kauf:1962}
\begin{equation}
\langle j^{\mathrm{(micro)\mu }}\rangle =j^{\mathrm{(free)\mu }}-cP^{\mu \nu
}{}_{,\nu }  \label{av-current}
\end{equation}
where the polarization tensor $P^{\mu \nu }$ is defined as an average of the
quadruple polarization moment of the molecules $p_{a}^{\mu \nu }$ (see
\cite{Kauf:1962}, \cite{Szek:1971} for details)
\begin{equation}
P^{\mu \nu }=\langle \sum_{a}\int d\tau _{a}p_{a}^{\mu \nu }\delta
^{4}(x-y_{a})\rangle .  \label{polarization-em}
\end{equation}
Then equations (\ref{av-bianchi}) can be rewritten as the macroscopic
equations\footnote{Gravitational macroscopic equations similar to (\ref{av-bianchi})
are known to have been proposed first in \cite{Bel:1961}.}
\begin{equation}
E_{\mu \nu \rho \sigma }{}^{,\sigma }=\kappa J_{\mu \nu \rho }^{\mathrm{
(free)}}  \label{av-bianchi-2}
\end{equation}
for the gravitational induction tensor $E_{\mu \nu \rho \sigma }$ defined as
\begin{equation}
E_{\mu \nu \rho \sigma }=\langle C_{\mu \nu \rho \sigma }\rangle +\kappa
P_{\mu \nu \rho \sigma }.  \label{gr-induction}
\end{equation}
The macroscopic equations (\ref{av-bianchi-2}) are analogous to the Maxwell
macroscopic equations obtainable by means of averaging the microscopic
equations (\ref{maxwell}) with $j^{\mathrm{(micro)}\mu }$ with taking into
account (\ref{av-current})
\begin{equation}
H^{\mu \nu }{}_{,\nu }=\frac{4\pi }{c}J^{\mathrm{(free)\mu }}
\label{av-maxwell}
\end{equation}
for the electromagnetic induction tensor $H_{\mu \nu }$ defined as
\begin{equation}
H^{\mu \nu }=\langle f^{\mu \nu }\rangle +4\pi P_{\mu \nu }.
\label{em-induction}
\end{equation}
Unfortunately, at this point the analogy between the electromagnetism and
gravitation which holds on the level of (\ref{bianchi}),
(\ref{matter-current-conserv}) and (\ref{maxwell}), (\ref{em-current-conserv})
breaks. Indeed, the formal similarity of (\ref{av-bianchi-2}),
(\ref{gr-induction}) and (\ref{av-maxwell}), (\ref{em-induction}) does not
possess the structural analogy between averaged electromagnetism and
gravitation: (A) the gravitational induction tensor $E_{\mu \nu \rho \sigma }
$ does not have any more the symmetries of the Weyl tensor compared with
$H_{\mu \nu }$ keeping the symmetries of $f_{\mu \nu }$; (B) it is
constructed from the second derivatives of the polarization tensor $Q_{\mu
\nu \rho \sigma }$ compared with the linear algebraic structure of the
electromagnetic induction tensor $H_{\mu \nu }$ in terms of the polarization
tensor $P_{\mu \nu }$ - it is thus impossible to proceed with the
formulation of phenomenological material relations between $E_{\mu \nu \rho
\sigma }$ and $\langle C_{\mu \nu \rho \sigma }\rangle $ as possible in
electromagnetism (relations between $H_{\mu \nu }$ and $\langle f_{\mu \nu
}\rangle $, or amongst the fields $\mathbf{E}$, $\mathbf{D}$, $\mathbf{B}$,
$\mathbf{H}$ and $\mathbf{J}$).

Even more important issue is that analysis of the macroscopic field equation
(\ref{av-bianchi})
\begin{equation}  \label{av-bianchi-orders}
{
\begin{array}[t]{c}
{\langle C_{\mu \nu \rho \sigma}{} \rangle}^{, \rho} \\
{\scriptscriptstyle {\mathcal{O}}(e)}
\end{array}
} = \kappa \langle J^{\mathrm{(micro)}}_{\mu \nu \rho} \rangle = {
\begin{array}[t]{c}
\kappa J^{\mathrm{(free)}}_{\mu \nu \rho} \\
{\scriptscriptstyle {\mathcal{O}}(1)}
\end{array}
} - {
\begin{array}[t]{c}
P_{\mu \nu \rho \sigma}{}^{, \sigma} \\
{\scriptscriptstyle {\mathcal{O}}(e^2)}
\end{array}
},
\end{equation}
where $e$ is a parameter measuring the value of deviation from the flat
space, requires one to put into agreement the orders of magnitude of all
quantities and reveals that the linearized Weyl tensor should be zero under
averaging $\langle C_{\mu \nu \rho \sigma} \rangle = 0$. The Weyl tensor
must be estimated in the perturbation theory up to the second order as it
was done for the matter current $J^{\mathrm{(micro)}}_{\mu \nu \rho}$ in the
left-hand side of (\ref{bianchi}). So considering some physical features of
the polarization tensor neither the macroscopic equations (\ref{av-bianchi}),
nor any other field equations had in fact been employed \cite{Szek:1971}.

On the basis of the expression
\begin{equation}
\langle t_{\mu \nu }^{\mathrm{(micro)}}\rangle =T_{\mu \nu }^{\mathrm{(free)}
}+\frac{1}{2}Q_{\mu \rho \nu \sigma }{}^{,\rho \sigma }
\label{av-energy-momentum}
\end{equation}
the following material relations have been suggested
\begin{equation}
Q_{i0j0}=\langle G_{ij}\rangle N=\epsilon _{g}C_{i0j0}  \label{gr-material}
\end{equation}
where $N$ is the average number of molecules per unit volume, $G_{ij}$ is
the quadrupole moment of a molecule
\begin{equation}
G_{ij}=\int \rho (x)\delta x_{i}\delta x_{j}d^{3}x,  \label{quadruple}
\end{equation}
$\rho =\rho (x)$ is the matter density in molecules, $\delta x_{i}$ is a
vector between neighboring particles of a molecule. The quantity $\epsilon
_{g}$ has been called the gravitational dielectric constant and in Newtonian
approximation found to be
\begin{equation}
\epsilon _{g}=\frac{1}{4}\frac{mA^{2}c^{2}}{\omega _{0}^{2}}N
\label{grav-dielectric-const}
\end{equation}
where $A$ is the average linear dimension of a typical molecule, $m$ is the
average mass of the molecules, $\omega _{0}^{2}$ is a typical frequency of
harmonically oscillating particles in molecules.

\section{The Macroscopic Gravity Equations}
\label{mgeqs}
\noindent
A macroscopic gravity approach in the perturbation theory up to the second
order on a particular background space-time taken to be a smooth weak
gravitational field is applied to write down a system of macroscopic field
equations: Isaacson's equations with a source incorporating the quadruple
gravitational polarization tensor, Isaacson's energy-momentum tensor of
gravitational waves and energy-momentum tensor of gravitational molecules
and corresponding equations of motion.

The gravitational field created by a number of particles represented by a
microscopic energy-momentum tensor $t_{\beta }^{\alpha \mathrm{(micro)}}$ is
defined by Einstein's equation
\begin{equation}
g^{\alpha \epsilon }r_{\epsilon \beta }-\frac{1}{2}\delta _{\beta }^{\alpha
}g^{\mu \nu }r_{\mu \nu }=-\kappa t_{\beta }^{\alpha \mathrm{(micro)}}
\label{ein}
\end{equation}
where $\kappa =8\pi G/c^{4}$ is Einstein's gravitational constant and $G$ is
Newton's gravitational constant. The Einstein equations (\ref{ein}) for the
microscopic distribution of gravitational molecules (\ref{molec}) have the
following form:
\begin{equation}
g^{\alpha \epsilon }r_{\epsilon \beta }-\frac{1}{2}\delta _{\beta }^{\alpha
}g^{\mu \nu }r_{\mu \nu }=-\kappa t_{\beta }^{\alpha \mathrm{(molec)}}.
\label{ein-molec}
\end{equation}

Averaging the left-hand side of the Einstein equations (\ref{ein-molec})
following the Isaacson's high-frequency approximation approach
\cite{Isaa:1968a}, \cite{Isaa:1968b}, using the averaging procedure
\cite{Zala:1992}$^{,}$\cite{Zala:1993} (one can also use Isaacson's averaging
procedure \cite{Isaa:1968a}, \cite{Isaa:1968b}, see also \cite{Zala:1996})
and with taking into account the expression \cite{Szek:1971} for the tensor
of gravitational quadrupole polarization $Q^{\mu \nu \rho \sigma }$ in terms
of the covariant gravitational quadrupole moment $q_{a}^{\mu \nu \rho \sigma
}$ (\ref{polar}) brings the averaged Einstein equations in the form:
\begin{equation}
R_{\mu \nu }^{(0)}-\frac{1}{2}g_{\mu \nu }^{(0)}R^{(0)}=-\kappa (T_{\mu \nu
}^{\mathrm{(free)}}+T_{\mu \nu }^{\mathrm{(GW)}}+\frac{1}{2}c^{2}Q_{\mu \rho
\nu \sigma }{}^{;\rho \sigma }),  \label{av-ein-molec}
\end{equation}
where $T_{\mu \nu }^{\mathrm{(GW)}}$ is Isaacson's energy-momentum tensor of
gravitational waves \cite{Isaa:1968a}, \cite{Isaa:1968b} and $T_{\mu \nu }^{
\mathrm{(free)}}$ is the energy-momentum tensor of molecules
\begin{equation}
T_{\mu \nu }^{\mathrm{(free)}}(x)=c^{-1}\sum_{a}\int m_{a}\frac{dy_{a}^{\mu }
}{d\tau _{i}}\frac{dy_{a}^{\nu }}{d\tau _{i}}\delta ^{4}[x-y_{a}^{\mu }(\tau
_{a})]d\tau _{a}.  \label{free}
\end{equation}
All members in equation (\ref{av-ein-molec}) can be shown to be of the same
order of magnitude ${\mathcal{O}}(1/L^{2})$. The macroscopic equations
(\ref{av-ein-molec}) give the equations of motion for
molecules\footnote{Under assumption that the background metric in the
left-hand side of (\ref{av-ein-molec}) represents a weak gravitational field on the flat background
one can use the covariant derivatives with respect the metric in all
relations instead of partial derivatives with respect to the flat metric.}
\begin{equation}
T^{\mathrm{(free)}\mu \nu }{}_{;\nu }=0,  \label{eq-motion-molec}
\end{equation}
conservation of the energy-momentum of gravitational waves
\begin{equation}
T^{\mathrm{(GW)}\mu \nu }{}_{;\nu }=0,  \label{conserv-gw}
\end{equation}
and an identity for the gravitational polarization
\begin{equation}
Q_{\mu \nu \rho \sigma }{}^{;\nu \sigma \mu }=0.  \label{ident-polar}
\end{equation}

The system of equations (\ref{av-ein-molec})-(\ref{ident-polar}) is
underdetermined because there are 20 unknown components of the tensor of
gravitational polarization. It is possible to formulate two natural material
relations. The first relation connects the traceless part of the quadrupole
polarization tensor
\begin{eqnarray}
\widetilde{Q}_{\mu \rho \nu \sigma } & = & Q_{\mu \rho \nu \sigma }+ \frac{1
}{2}(-g_{\mu \nu }P_{\rho \sigma }+g_{\mu \sigma }P_{\rho \nu }-g_{\rho
\sigma }P_{\mu \nu}+g_{\rho \nu }P_{\mu \sigma }) + \nonumber \\
&  & \frac{1}{2}S(g_{\mu \nu }g_{\rho \sigma }-g_{\mu \sigma
}g_{\rho \nu }), \label{polar-decomp}
\end{eqnarray}
where $P_{\rho \sigma }=Q^{\mu }{}_{\rho \mu \sigma }$, $S=P_{\rho }^{\rho }$
, with the traceless energy-momentum tensor of gravitational waves $T_{\mu
\nu }^{\mathrm{(GW)}}$
\begin{equation}
\frac{c^{2}}{2}\widetilde{Q}_{\mu \rho \nu \sigma }{}^{;\rho \sigma
}=\lambda T_{\mu \nu }^{\mathrm{(GW)}},  \label{material-1}
\end{equation}
where $\lambda =\lambda (x)$ the gravitational radiation polarization
factor. Relation (\ref{material-1}) can be shown to be always valid in the
geometrical optics limit.

The second material relation connects the remaining part of the polarization
tensor $Q_{\mu \rho \nu \sigma }$, namely its trace $P_{\rho \sigma }$, with
a projection of the curvature tensor on the world line of an observer (the
electric part of the curvature tensor)
\begin{equation}
P_{\rho \sigma }=\epsilon R_{\mu \rho \nu \sigma }^{(0)}u^{\mu }u^{\nu },
\label{material-2}
\end{equation}
where $u^{\mu }$ is the observer 4-velocity (4-velocity of the molecule
center of mass) and $\epsilon =\epsilon (x)$ is the macroscopic medium
polarization factor. The relation (\ref{material-2}) can be shown to lead to
the correct expression for the 3-tensor of the average quadrupole
gravitational moment \cite{Szek:1971} so that
\begin{equation}
P_{\mu \nu }=(P_{00}=0,P_{0i}=0,P_{ij}=\langle G_{ij}\rangle N).
\label{q-trace}
\end{equation}
where $N$ is the average number of molecules per unit volume, $\langle
G_{ij}\rangle $ is the averaged quadrupole moment of a molecule
(\ref{quadruple}). Then the material relation \cite{Szek:1971} can be recovered
in the form
\begin{equation}
Q_{i0j0}=\langle G_{ij}\rangle N=\epsilon _{g}R_{i0j0}
\label{gr-material-improv}
\end{equation}
that gives $\epsilon =\epsilon _{g}$ with the gravitational dielectric
constant $\epsilon _{g}$ defined in \cite{Szek:1971} as
(\ref{grav-dielectric-const}).

Thus the system of equations (\ref{av-ein-molec})-(\ref{conserv-gw}),
(\ref{material-1}), (\ref{material-2}) is fully determined and can be used to
find the gravitational and polarization fields for the macroscopic
gravitating systems.

\section{The Static Weak-field Macroscopic Medium}
\label{medium}
\noindent
The averaged Einstein equations (\ref{av-ein-molec}) have been derived under
assumption of the weak gravitational field though they can be considered to
be formally valid for any background metric $g_{\mu \nu }^{(0)}$ with a
given tensor of gravitational quadrupole polarization $Q^{\mu \nu \rho
\sigma }$ and the material relations (\ref{material-1}) and (\ref{material-2}
). The definition of the tensor of gravitational quadrupole polarization $
Q^{\mu \nu \rho \sigma }$ (\ref{polar}) adopted here is essentially valid on
the flat space-time background due to the used definitions of molecule's
center of mass \cite{Szek:1971}, \cite{Syng:1956}. Therefore, the averaged
Einstein equations (\ref{av-ein-molec}) with (\ref{polar}) can be only
consistently applied in the framework of the weak-field approximation.

A model of static weak-field macroscopic medium with quadrupole
gravitational polarization is considered here. The model is based on three
assumptions.

(1) Newtonian gravity conditions for the energy-momentum tensor of molecules
$T_{\mu \nu }^{\mathrm{(free)}}$,
\begin{equation}
T_{00}^{\mathrm{(free)}}\gg T_{ij}^{\mathrm{(free)}},\quad T_{00}^{\mathrm{\
(free)}}\gg T_{0i}^{\mathrm{(free)}},\quad T_{00}^{\mathrm{(free)}}=T_{\mu
\nu }^{\mathrm{(free)}}u^{\mu }u^{\nu }=\mu c^{2},  \label{newton}
\end{equation}
where the observer 4-velocity (4-velocity of the molecule center of mass) $
u^{\nu }$ is $u^{\nu }=(1,0,0,0)$ and $\mu $ is the macroscopic matter
density. The condition (\ref{newton}) means that the gravitational field
created by gravitational molecules is essentially Newtonian.

(2) The macroscopic metric tensor $g_{\mu \nu }^{(0)}$ is static,
\begin{equation}
\frac{\partial g_{\mu \nu }^{(0)}}{\partial t}=0,  \label{staticity}
\end{equation}
which means that is the macroscopic matter density $\mu $ depends only on
spatial coordinates, $\mu =\mu (x^{a})$.

(3) The condition of the weak field approximation for the macroscopic metric
tensor $g_{\mu \nu }^{(0)}$,
\begin{equation}
g_{\mu \nu }^{(0)}=\eta _{\mu \nu }+eh_{\mu \nu },  \label{weak}
\end{equation}
where is $\eta _{\mu \nu }$ the flat space-time metric,
$\eta _{0 0}=-1$, $\eta _{1 1}=1$, $\eta _{2 2}=1$, $\eta _{3 3}=1$ and $\eta
_{\mu \nu }=0$ if $\mu \neq \nu $, $h_{\mu \nu }$ is the arbitrary
perturbation functions depending here only on spatial coordinates and $e$ is
the smallness parameter, $e\ll 1$.

In general relativity the conditions (\ref{newton})-(\ref{weak}) for a
microscopic energy-momentum tensor $t_{\beta }^{\alpha \mathrm{(micro)}}$
and for the metric tensor $g_{00}=-\left( 1+\frac{2\varphi }{c^{2}}\right) $
, $g_{0i}=0$, $g_{ij}=0$, are known to lead to the Newtonian limit of the
Einstein equations (\ref{ein}) to result in the Poisson equation for the
Newtonian gravitational potential $\varphi =\varphi (x^{a})$
\begin{equation}
\Delta \varphi =4\pi G\mu .  \label{poisson}
\end{equation}

The Newtonian limit of the macroscopic gravity equations (\ref{av-ein-molec}
) under conditions (\ref{newton})-(\ref{weak}) for the macroscopic tensor $
g_{\mu \nu }^{(0)}$
\begin{equation}
g_{0 0 }^{(0)}=-\left( 1+\frac{2\varphi }{c^{2}}\right), \quad g_{1 1}^{(0)}=1,
\quad g_{2 2}^{(0)}=1, \quad g_{3 3}^{(0)}=1, \quad g_{\mu \nu }^{(0)}=0, \quad \mu \neq \nu ,
\label{potential}
\end{equation}
should bring a generalization of the Poisson equation which incorporates the
effect of gravitational quadrupole polarization.

For the case of static weak-field macroscopic medium Isaacson's
energy-momentum tensor of gravitational waves $T_{\mu \nu }^{\mathrm{(GW)}}$
vanishes
\begin{equation}
T_{\mu \nu }^{\mathrm{(GW)}}=0,  \label{isaacson's}
\end{equation}
and no gravitational radiation polarization factor $\lambda $ is involved
\begin{equation}
\lambda =0.  \label{lambda}
\end{equation}
The the macroscopic medium polarization factor $\epsilon $ is here
\begin{equation}
\epsilon =\epsilon _{g}  \label{epsilon}
\end{equation}
with the gravitational dielectric constant $\epsilon _{g}$ defined
\cite{Szek:1971} as (\ref{grav-dielectric-const}).\

\section{The Equation for the Macroscopic Gravitational Potential}
\label{potential_eq}
\noindent
Calculation of the equation for the macroscopic gravitational potential $
\varphi $ from the macroscopic gravity equations (\ref{av-ein-molec}) under
conditions (\ref{newton})-(\ref{weak}) for the macroscopic tensor $g_{\mu
\nu }^{(0)}$ (\ref{potential}) brings the equation
\begin{equation}
\Delta \varphi =4\pi G\mu +\frac{4\pi G\epsilon _{g}}{3c^{2}}\Delta
^{2}\varphi  \label{macropotential}
\end{equation}
where $\Delta ^{2}\varphi \equiv \Delta (\Delta \varphi )$ is the Laplacian
of the Laplacian of $\varphi $. This is a non-trivial generalization of the
Poisson equation for the gravitational potential $\varphi $ of Newtonian
gravity (\ref{poisson}). This is a biharmonic equation due the presence of
the term $\Delta ^{2}\varphi $. The equation (\ref{macropotential}) involves
a singular perturbation, since in case of the vanishing gravitational
dielectric constant, $\epsilon _{g}=0$, this equation becomes the Poisson
equation, but if $\epsilon _{g}\neq 0$, this equations change its operator
structure to be of the fourth order equation in partial derivatives of $
\varphi $ as compared with the Poisson second order partial differential
equation.

It is convenient to introduce the factor
\begin{equation}
\frac{1}{k^{2}}=\frac{4\pi G\epsilon _{g}}{3c^{2}}  \label{k2}
\end{equation}
with $k$ having a physical dimension of inverse length, $\left[ k^{-2}\right]
=\mathrm{length}^{2}$. Then the equation (\ref{macropotential}) takes the
form
\begin{equation}
\Delta \varphi =4\pi G\mu +\frac{1}{k^{2}}\Delta ^{2}\varphi .
\label{macropotential_k}
\end{equation}
By using the definitions of the gravitational dielectric constant
$\epsilon _{g}$ (\ref{grav-dielectric-const}), the characteristic
oscillation frequency of molecule's constituents $\omega
_{0}^{2}$, macroscopic matter density $\mu =3m/4\pi A^{3}$ and the
average number of molecules per unit volume $N=4\pi D^{3}/3$ with
$D$ as a mean distance between molecules, the factor $k^{-2}$ can
be shown to have the following form
\begin{equation}
\frac{1}{k^{2}}=\frac{1}{4\theta }\left( \frac{A^{3}}{D^{3}}\right) A^{2}.
\label{k2A2}
\end{equation}
Here the dimensionless factor $\theta $,
\begin{equation}
\theta =\frac{\omega _{0}^{2}}{4\pi G\mu /3},  \label{theta}
\end{equation}
reflects the nature of field responsible for bounding of discrete matter
constituents into molecules. If $\theta \approx 1$, the molecules of
self-gravitating macroscopic medium are considered to be gravitationally
bound. For instance, considering a macroscopic model of galaxy as a
self-gravitating macroscopic medium consisting of gravitational molecules
taken as double stars, $\theta \approx 1$ as such galactic molecules are
gravitationally bound. If one takes the molecules to be of electron-proton
type, like atoms, the factor $\theta \approx 10^{40}$, which makes the
factor $k^{-2}$ essentially insignificant.

The dimensionless ratio $A/D$ reflects the structure of macroscopic medium.
If $(A/D)\approx 1$, the macroscopic medium behaves itself like a liquid or
solid. If $(A/D)<1$, the macroscopic medium behaves itself like a gas. For
the macroscopic galactic model for the present epoch the macroscopic medium
is like a gas, since $(A/D)\approx 10^{-1}-10^{-2}$, which makes the factor $
A^{3}/D^{3}$ to be of order of $10^{-3}-10^{-6}$. However, for earlier times
of galaxy formulation this factor can be expected to be of much greater
order of magnitude up to $1-10$.

It is useful to introduce a dimensionless factor $k^{-2}L^{-2}$,
\begin{equation}
\frac{1}{k^{2}L^{2}}=\frac{1}{4\theta }\left( \frac{A^{3}}{D^{3}}\right)
\left( \frac{A^{2}}{L^{2}}\right) ,  \label{k2L2}
\end{equation}
where $L$ is the characteristic scale of the macroscopic gravitational
field. The dimensionless ratio $A/L$ reflects the scale of significant
change in the macroscopic gravitational field. If $(A/L)\approx 1$, the
macroscopic gravitational field changes significantly on the scale $L$. If $
(A/D)\ll 1$, the scale $L$ does not reflect the presence of gravitational
molecules.

Thus, the structure of the factor $k^{-2}$ in (\ref{k2A2}), or $k^{-2}L^{-2}$
in (\ref{k2L2}), is model dependent, and its particular value is fully
determined by a particular model of self-gravitating macroscopic matter.
When a macroscopic medium has the factor
\begin{equation}
\frac{1}{k^{2}L^{2}}=\frac{1}{4\theta }\left( \frac{A^{3}}{D^{3}}\right)
\left( \frac{A^{2}}{L^{2}}\right) \approx 1,  \label{k2L2=1}
\end{equation}
the equation for the macroscopic gravitational potential $\varphi $ is
(\ref{macropotential_k}) which is convenient to write in the following form
\begin{equation}
\Delta ^{2}\varphi -k^{2}\Delta \varphi =-4\pi Gk^{2}\mu .
\label{macropotential_k2}
\end{equation}
It can be rewritten as a system of two second order partial differential
equations
\begin{equation}
\Delta \varphi =f,  \label{L1-1}
\end{equation}
\begin{equation}
\Delta f-k^{2}f=-4\pi Gk^{2}\mu ,  \label{L1-2}
\end{equation}
for the unknowns $\varphi (x,y,z)$ and $f(x,y,z)$ with given $\mu (x,y,z)$
and $k^{2}$. In this case of the significant gravitational quadrupole
polarization of a macroscopic medium, which will be referred to as Case I,
the equations (\ref{L1-2}) is singular with respect to the gravitational
dielectric constant $\epsilon _{g}$ because $k^{2}\sim 1/\epsilon _{g}$ and
the limit $\epsilon _{g}\rightarrow 0$ cannot be accomplished in a solution
to (\ref{L1-1}) and (\ref{L1-2}). Equations (\ref{L1-1}) and (\ref{L1-2}) do
not have a limit to the Poisson equation (\ref{poisson}) of Newtonian
gravity, nor does a solution to (\ref{L1-1}) and (\ref{L1-2}) have a limit
to a solution to (\ref{poisson}). It should be pointed out that the
equations ( \ref{L1-1}) and (\ref{L1-2}) are essentially valid either inside
the macroscopic matter source, or outside the matter source in the close
vicinity of the macroscopic matter configuration boundary where the effect
of strong gravitational quadrupole polarization is still significant,
\begin{equation}
\frac{1}{k^{2}}\Delta ^{2}\varphi \approx \frac{1}{k^{2}L^{2}}\Delta \varphi
\gg \Delta \varphi .  \label{L1-3}
\end{equation}
As far as asymptotically $\Delta \varphi /k^{2}L^{2}$ becomes much less than
$\Delta \varphi $ the equations (\ref{L1-1}) and (\ref{L1-2}) are not valid
anymore. See the general solution to equations (\ref{L1-1}) and (\ref{L1-2})
and their asymptotic behavior in Section \ref{case_I}.

When a macroscopic medium has the factor
\begin{equation}
\frac{1}{k^{2}L^{2}}=\frac{1}{4\theta }\left( \frac{A^{3}}{D^{3}}\right)
\left( \frac{A^{2}}{L^{2}}\right) \ll 1,  \label{k2L2<1}
\end{equation}
the equation (\ref{macropotential_k}) for the macroscopic gravitational
potential $\psi $ can be rewritten effectively in the form of the Poisson
equation
\begin{equation}
\Delta \psi =4\pi G\mu \left( 1+\frac{1}{k^{2}L^{2}}\right) ,  \label{L2}
\end{equation}
for the unknowns $\psi (x,y,z)$ with given $\mu (x,y,z)$ and $k^{2}L^{2}$.
It is possible because the $\Delta ^{2}\psi $ term in (\ref{macropotential_k}
) is well approximated as
\begin{equation}
\frac{1}{k^{2}}\Delta ^{2}\psi \approx \frac{1}{k^{2}L^{2}}\Delta \psi \ll
\Delta \psi .  \label{L2-2}
\end{equation}
In this case of the insignificant gravitational quadrupole polarization of a
macroscopic medium, which will be referred to as Case II, the equation
(\ref{L2}) is nonsingular with respect to the gravitational dielectric constant $
\epsilon _{g}$ because $1/k^{2}\sim \epsilon _{g}$ and the limit $\epsilon
_{g}\rightarrow 0$ can be accomplished in the solution to (\ref{L2}).
Equation (\ref{L2}) does have a limit to the Poisson equation
(\ref{poisson}) of Newtonian gravity, and a solution to (\ref{L2}) always has a limit to a
solution to (\ref{poisson}). It should be pointed out that the equation
(\ref{L2}) holds either inside the macroscopic matter source, or outside the
matter source. A solution to equation (\ref{L2}) gives the macroscopic
gravitational potential $\psi (x,y,z)$ defined everywhere.

The equation (\ref{L2}) describes also asymptotic behavior, $
L^{2}\rightarrow \infty $, for the field of the macroscopic gravitational
potential $\varphi (x,y,z)$for a macroscopic matter distribution with strong
gravitational quadrupole polarization. A solution to equations (\ref{L1-1})
and (\ref{L1-2}) must be matched to a solution to equation (\ref{L2}) on a
surface outside the source between the source boundary and the nearby zone
to get the macroscopic gravitational potential $\varphi (x,y,z)$ in
asymptotic region and, as a result, to define it everywhere.

The Cases I and II have different equations and different physics, which is
reflected by solutions to equations (\ref{L1-1}) and (\ref{L1-2}) and
equation (\ref{L2}).

\section{Case I: The Strong Quadrupole Polarization}
\label{case_I}
\noindent
The system of equations (\ref{L1-1}) and (\ref{L1-2}) can be solved by
separation of variables. Consider the equation (\ref{L1-2}) in the spherical
coordinates $(x,y,z)\rightarrow (r,\theta ,\phi )$ when the Laplacian
becomes
\begin{equation}
\Delta =\frac{1}{r^{2}}\frac{\partial }{\partial r}\left( r^{2}\frac{
\partial }{\partial r}\right) -\frac{\hat{L}^{2}}{r^{2}},\quad \hat{L}^{2}=
\frac{\partial ^{2}}{\partial \theta ^{2}}+\coth \frac{\partial }{\partial
\theta }+\frac{1}{\sin ^{2}\theta }\frac{\partial ^{2}}{\partial \phi ^{2}}.
\label{laplacian}
\end{equation}
Applying the method of separation of variables for the function $f(r,\theta
,\phi )$ in (\ref{L1-2}) with assuming the angular part is given by the
spherical harmonics $Y_{m}^{l}(\theta ,\phi )$ \cite{Erde-etal:1953},
\begin{equation}
f(r,\theta ,\phi )=\frac{S(r)}{r}Y_{m}^{l}(\theta ,\phi ),
\label{separation}
\end{equation}
one can represent the Green function in a spherical harmonics expansion
\cite{Arfk:1985} as
\begin{equation}
G(r_{1},r_{2})=\sum_{l=0}^{\infty
}\sum_{m=-l}^{l}g_{l}(r_{1},r_{2})Y_{m}^{l}(\theta _{1},\phi
_{1})Y_{m}^{l}(\theta _{2},\phi _{2}),  \label{green(sh)}
\end{equation}
which gives the equation for the radial Green function $g_{l}(r_{1},r_{2})$
\begin{equation}
r_{1}\frac{d^{2}S}{dr_{1}^{2}}
[r_{1}g_{l}(r_{1},r_{2})]-k^{2}r^{2}g_{l}(r_{1},r_{2})-l(l+1)g_{l}(r_{1},r_{2})=-4\pi \delta (r_{1}-r_{2}).
\label{green(r)}
\end{equation}
It has the solution \cite{Arfk:1985}
\begin{equation}
g_{l}(r_{1},r_{2})=ki_{l}(kr_{<})k_{l}(kr_{>}),\quad
i_{l}(kr),~r_{1}<r_{2},\quad k_{l}(kr),~r_{1}>r_{2}  \label{green(r)_sol}
\end{equation}
and the solution for $f(r,\theta ,\phi )$ for $r_{1}>r_{2}$ has the form
\begin{eqnarray}
f({{\mathbf{r}}_{1}}) & = & Gk^{2}\sum_{l=0}^{\infty
}\sum_{m=-l}^{l}ki_{l}(kr_{1})Y_{m}^{l}(\theta _{1},\phi _{1})\times \nonumber \\
&  & \int \mu ({{\mathbf{r}}_{2}})k_{l}(kr_{2})Y_{m}^{l}(\theta _{2},
\phi_{2})d\phi _{2}\sin \theta _{2}d\theta _{2}r_{2}^{2}dr_{2}. \label{f(r)}
\end{eqnarray}
This is a multipole expansion of $f({\mathbf{r}}_{1})$ where the particular
structure of the function of a macroscopic matter distribution
$\mu ({\mathbf{r}}_{2})$ will make the particular structure of multipole expansion.

Now one can solve the Poisson equation (\ref{L1-1}) with known
$f(\mathbf{r})$ to obtain $\varphi (\mathbf{r})$. But rather than continuing solving the
system in spherical harmonic representation, let us solve the system in
general form by using Green's function method. The solutions are very useful
for illustration of the character of the macroscopic gravitational potential
$\varphi (\mathbf{r})$ for the static weak-field approximation with the
quadrupole moment tensor. One can always accomplish expansion of the general
solution with respect to the spherical harmonics to get a multipole
expansion.

The system of equations (\ref{L1-1}) and (\ref{L1-2}) must be supplemented
by boundary conditions for the unknown $f(\mathbf{r})$ and
$\varphi (\mathbf{r})$ with given $\mu (\mathbf{r})$ and $k^{2}$:
\begin{equation}
\lim_{\left\vert {\mathbf{r}}\right\vert \rightarrow 0^{+}}\varphi ({\mathbf{r}})
={\mathrm{exists~and~is~bounded,}}\quad \lim_{\left\vert {\mathbf{r}}
\right\vert \rightarrow \infty }\varphi ({\mathbf{r}})=0,\quad 0<\left\vert
{\mathbf{r}}\right\vert <\infty ,  \label{phi_bc}
\end{equation}
\begin{equation}
\lim_{\left\vert {\mathbf{r}}\right\vert \rightarrow 0^{+}}f({\mathbf{r}})=
{\mathrm{exists~and~is~bounded,}}\quad \lim_{\left\vert {\mathbf{r}}\right\vert
\rightarrow \infty }f({\mathbf{r}})=0,\quad 0<\left\vert {\mathbf{r}}\right\vert
<\infty .  \label{f_bc}
\end{equation}
The system of partial differential equations (\ref{L1-1}) and (\ref{L1-2})
with the boundary conditions (\ref{phi_bc}) and (\ref{f_bc}) is a Dirichlet
boundary value problem on interval $0<\left\vert \mathbf{r}\right\vert
<\infty $. One can consider also two Dirichlet problems for intervals $
(0,\left\vert {{\mathbf{r}}}_{0}\right\vert )$ and $(\left\vert {{\mathbf{r}}}
_{0}\right\vert ,\infty )$ where ${{\mathbf{r}}}_{0}$ is the radius vector of a
macroscopic matter configuration.

To solve first the nonhomogeneous modified Helmholtz equation (\ref{L1-2}),
one needs to find the corresponding Green function
\begin{equation}
\Delta G_{f}({\mathbf{r}}_{1}{\mathbf{,r}}_{2})-k^{2}G_{f}({\mathbf{r}}_{1},
{\mathbf{r}}_{2})=-4\pi \delta ({\mathbf{r}}_{1}-{{\mathbf{r}}_{2}})  \label{G(f)}
\end{equation}
with the boundary condition
\begin{equation}
\lim_{\left\vert {\mathbf{r}}_{1}\right\vert \rightarrow \infty }
G_{f}({\mathbf{r}}_{1}{\mathbf{,r}}_{2})=0.  \label{G(f)_bc}
\end{equation}
The Green function can be found \cite{Arfk:1985} to be
\begin{equation}
G_{f}({\mathbf{r}}_{1}{\mathbf{,r}}_{2})=\frac{e^{-k\left\vert {\mathbf{r}}_{1}-
{\mathbf{r}}_{2}\right\vert }}{\left\vert {\mathbf{r}}_{1}-
{{\mathbf{r}}}_{2}\right\vert }  \label{G(f)_exp}
\end{equation}
and the solution for $f({\mathbf{r}})$ is
\begin{equation}
f({\mathbf{r}}_{1})=Gk^{2}\int \frac{e^{-k\left\vert {\mathbf{r}}_{1}-{\mathbf{r}}
_{2}\right\vert }}{\left\vert {\mathbf{r}}_{1}-{\mathbf{r}}_{2}\right\vert }\mu
( {\mathbf{r}}_{2})dV_{2}.  \label{f(r)_exp}
\end{equation}
The second equation (\ref{L1-1}) has the Green function
\begin{equation}
G_{\varphi }({\mathbf{r}}_{1}{\mathbf{,r}}_{2})=\frac{1}{\left\vert {\mathbf{r}}
_{1}-{\mathbf{r}}_{2}\right\vert }  \label{G(phi)_exp}
\end{equation}
as the general solution to the Green equation,
\begin{equation}
\Delta G_{\varphi }({\mathbf{r}}_{1}{\mathbf{,r}}_{2})=-4\pi \delta ({\mathbf{r}}
_{1}-{\mathbf{r}}_{2}),  \label{G(phi)}
\end{equation}
to bring the solution for $\varphi (\mathbf{r})$
\begin{equation}
\varphi ({\mathbf{r}}_{1})=-\frac{1}{4\pi }\int \frac{f({\mathbf{r}}_{2})}{
\left\vert {\mathbf{r}}_{1}-{\mathbf{r}}_{2}\right\vert }dV_{2}.
\label{phi(r)_exp}
\end{equation}
Now the solution for the equation (\ref{macropotential_k2}) can be written
as
\begin{equation}
\varphi ({\mathbf{r}}_{1})=-\frac{Gk^{2}}{4\pi }\int \frac{dV_{2}}{\left\vert
{\mathbf{r}}_{1}-{\mathbf{r}}_{2}\right\vert }\int \frac{dV_{3}e^{-k\left\vert
{\mathbf{r}}_{2}-{\mathbf{r}}_{3}\right\vert }}{\left\vert {\mathbf{r}}_{2}-
{\mathbf{r}}_{3}\right\vert }\mu ({\mathbf{r}}_{3}).  \label{phi(r)_tot}
\end{equation}
If the macroscopic gravitational potential $\varphi (\mathbf{r})$ is
calculated in the origin of the coordinate system, that is
${\mathbf{r}}_{1}=0$, ${\mathbf{r}}_{2}=\mathbf{R}$, the formula (\ref{phi(r)_tot}) becomes
\begin{equation}
\varphi =-\frac{Gk^{2}}{4\pi }\int \frac{dV_{R}}{\left\vert \mathbf{R}
\right\vert }\int \frac{dV_{r}e^{-k\left\vert \mathbf{R}-\mathbf{r}
\right\vert }}{\left\vert \mathbf{R}-\mathbf{r}\right\vert }\mu (\mathbf{r}).
\label{phi(R)_tot}
\end{equation}

The asymptotic form of the solution (\ref{phi(R)_tot}) as $\left\vert
\mathbf{R}\right\vert \rightarrow 0^{+}$ is given by
\begin{equation}
\lim_{\left\vert \mathbf{r}\right\vert \rightarrow 0^{+}}\varphi \simeq
-4\pi Gk^{2}\mu (\mathbf{r})\left\vert {\mathbf{R}}\right\vert ^{4}.
\label{phi(0)}
\end{equation}
It shows that the macroscopic gravitational potential $\varphi (\mathbf{r})$
has better analytical properties as $\left\vert \mathbf{r}\right\vert
\rightarrow 0^{+}$ then the gravitational potential of the Poisson equation.

The asymptotic form of the solution (\ref{phi(R)_tot}) as $\left\vert
\mathbf{R}\right\vert \rightarrow \infty $ is given by
\begin{equation}
\lim_{\left\vert \mathbf{r}\right\vert \rightarrow \infty }\varphi \simeq
-Gk^{2}\frac{Q_{ij}R_{i}R_{j}e^{-k\left\vert \mathbf{R}\right\vert }}{{\
\left\vert \mathbf{R}\right\vert }^{3}}  \label{phi(inf)}
\end{equation}
where $Q_{ij}$ is the total quadrupole moment of the macroscopic mass
distribution
\begin{equation}
Q_{ij}=\int \mu ({\mathbf{r}})r_{i}r_{j}dV_{r}.  \label{Q}
\end{equation}
The formula (\ref{phi(inf)}) shows that the macroscopic gravitational
potential $\varphi (\mathbf{r})$ fades out in the vicinity of the
macroscopic matter source boundary on the characteristic distance $k^{-1}$.
For the distances $\left\vert \mathbf{R}\right\vert \gg k^{-1}$ the
inequality (\ref{L1-3}) does not hold and the inequality (\ref{L2-2}) is
valid. Therefore to find a proper asymptotic form of the macroscopic
gravitational potential $\varphi (\mathbf{r})$ for $\left\vert \mathbf{R}
\right\vert \gg k^{-1}$ one should solve the equation (\ref{L2}) and match
the solution (\ref{psi(r)_exp} ), see Section \ref{case_II}, with the
solution (\ref{phi(R)_tot}).

\section{Case II: The Weak Quadrupole Polarization}
\label{case_II}
\noindent
The equation (\ref{L2}) can be solved by the Green function method. To solve
the Poisson equation (\ref{L2}), one needs to find the corresponding Green
function
\begin{equation}
\Delta G_{\psi }({\mathbf{r}}_{1}{\mathbf{,r}}_{2})=-4\pi \delta
({\mathbf{r}}_{1}- {\mathbf{r}}_{2})  \label{G(psi)}
\end{equation}
with the boundary condition
\begin{equation}
\lim_{\left\vert {\mathbf{r}}_{1}\right\vert \rightarrow \infty }G_{\psi }(
{\mathbf{r}}_{1}{\mathbf{,r}}_{2})=0.  \label{G(psi)_bc}
\end{equation}
The Green function can be found \cite{Arfk:1985} to be
\begin{equation}
G_{\psi }({\mathbf{r}}_{1}{\mathbf{,r}}_{2})=\frac{1}{\left\vert {\mathbf{r}}_{1}-
{\mathbf{r}}_{2}\right\vert }  \label{G(psi)_exp}
\end{equation}
and the solution for $\psi (\mathbf{r})$ is
\begin{equation}
\psi ({\mathbf{r}}_{1})=-G\left( 1+\frac{1}{k^{2}L^{2}}\right) \int \frac{\mu
({\mathbf{r}}_{2})}{\left\vert {\mathbf{r}}_{1}-{\mathbf{r}}_{2}\right\vert }
dV_{2}.  \label{psi(r)_exp}
\end{equation}
If the macroscopic gravitational potential $\psi (\mathbf{r})$ is calculated
in the origin of the coordinate system, that is $\mathbf{r}_{1}=0$, $\mathbf{
\ \ r}_{2}=\mathbf{R}$, the formula (\ref{phi(r)_tot}) becomes
\begin{equation}
\psi =-G\left( 1+\frac{1}{k^{2}L^{2}}\right) \int \frac{\mu (\mathbf{r})}{
\left\vert \mathbf{R}-\mathbf{r}\right\vert }dV_{r}.  \label{psi(R)_exp}
\end{equation}

Behavior near the center and at infinity are essentially the same up to
renormalization by the factor $\left( 1+1/k^{2}L^{2}\right) $. The multipole
expansion \cite{Land-Lifs:1975} for the macroscopic gravitational potential $
\psi $ with $L\sim R_{0}$ has the form
\begin{equation}
\psi =-G\left( 1+\frac{1}{k^{2}L^{2}}\right) \left\{ \frac{M}{R_{0}}+\frac{1
}{6}D_{ij}\frac{\partial }{\partial X_{i}}\frac{\partial }{\partial X_{j}}
\frac{1}{R_{0}}+...\right\} .  \label{psi(R)_multi}
\end{equation}
The factor $\left( 1+1/k^{2}L^{2}\right) $ renormalizes the whole
potential to contribute into all terms in the multipole expansion.

\section*{Acknowledgements}
\noindent
Roustam Zalaletdinov would like to thank Remo Ruffini for hospitality
and hosting his NATO-CNR Fellowship in ICRA.

\end{document}